\documentclass[graybox]{svmult}

\usepackage{mathptmx}       
\usepackage{helvet}         
\usepackage{courier}        
\usepackage{type1cm}        
\usepackage{makeidx}         
\usepackage{graphicx}        
\usepackage{multicol}        
\usepackage[bottom]{footmisc}
\usepackage{url}
\usepackage{sidecap}
\usepackage{fancyhdr}

\makeindex             

\begin{document}

\title*{A High-Performance Interactive Computing Framework for Engineering Applications}
\author{Jovana Kne\v{z}evi\'{c}, Ralf-Peter Mundani, Ernst Rank}
\institute{Jovana Kne\v{z}evi\'{c} \at Technische Universit\"{a}t M\"{u}nchen, Arcisstr. 21, 80333 M\"{u}nchen \email{knezevic@bv.tum.de} \and Ralf-Peter Mundani \at Technische Universit\"{a}t M\"{u}nchen, Arcisstr. 21, 80333 M\"{u}nchen \email{mundani@tum.de} \and Ernst Rank \at Technische Universit\"{a}t M\"{u}nchen, Arcisstr. 21, 80333 M\"{u}nchen \email{ernst.rank@tum.de}}
\maketitle

\pagestyle{empty}
\thispagestyle{fancy}
\lhead{}
\chead{}
\rhead{}
\lfoot{\scriptsize This is a pre-print of an article published in Bader~M., Bungartz~HJ., Weinzierl~T.\ (eds) Advanced Computing. Lecture Notes in Computational Science and Engineering, vol 93, 2013. The final authenticated version is available online at: https://doi.org/10.1007/978-3-642-38762-3\_9}
\cfoot{}
\rfoot{}

\flushbottom

\abstract{To harness the potential of advanced computing technologies,
efficient (real time) analysis of large amounts of data is as essential as are
front-line simulations. In order to optimise this process, experts need to be
supported by appropriate tools that allow to interactively guide both the
computation and data exploration of the underlying simulation code. The main
challenge is to seamlessly feed the user requirements back into the simulation.
State-of-the-art attempts to achieve this, have resulted in the insertion of
so-called check- and break-points at fixed places in the code. Depending on the
size of the problem, this can still compromise the benefits of such an attempt,
thus, preventing the experience of real interactive computing. To leverage the
concept for a broader scope of applications, it is essential that a user
receives an immediate response from the simulation to his or her changes. Our
generic integration framework, targeted to the needs of the computational
engineering domain, supports distributed computations as well as on-the-fly
visualisation in order to reduce latency and enable a high degree of
interactivity with only minor code modifications. Namely, the regular course of
the simulation coupled to our framework is interrupted in small, cyclic
intervals followed by a check for updates. When new data is received, the
simulation restarts automatically with the updated settings (boundary
conditions, simulation parameters, etc.). To obtain rapid, albeit approximate
feedback from the simulation in case of perpetual user interaction, a
multi-hierarchical approach is advantageous. Within several different
engineering test cases, we will demonstrate the flexibility and the
effectiveness of our approach.}

\section{Introduction}
\label{sec:1}
Simulation of very complex physical phenomena becomes a realistic endeavour
with the latest advances in hardware technologies, sophisticated (numerical)
algorithms, and efficient parallelisation strategies. It consists of modelling
a domain of a physical problem, applying appropriate boundary conditions, and
doing numerical approximation for the governing equations, often with a linear
or non-linear system as outcome. When the system is solved, the result is
validated and visualised for more intuitive interpretations.

All the aforementioned cycles -- pre-processing, computation, and
post-processing -- can be very time consuming, depending on the discretisation
parameters, e.\,g., and moreover, are traditionally carried out as a sequence
of steps. The ever-increasing range of specialists in developing engineering
fields has necessitated an interactive approach with the computational model.
This requires real-time feedback from the simulation during program runtime,
while experimenting with different simulation setups. For example, the geometry
of the simulated scene can be modified interactively altogether with boundary
conditions or a distinct feature of the application, thus, the user can gain
``insight concerning parameters, algorithmic behaviour, and optimisation
potentials''~\cite{Mulder.1999}.

\begin{figure}[h]
\includegraphics[scale=0.5]{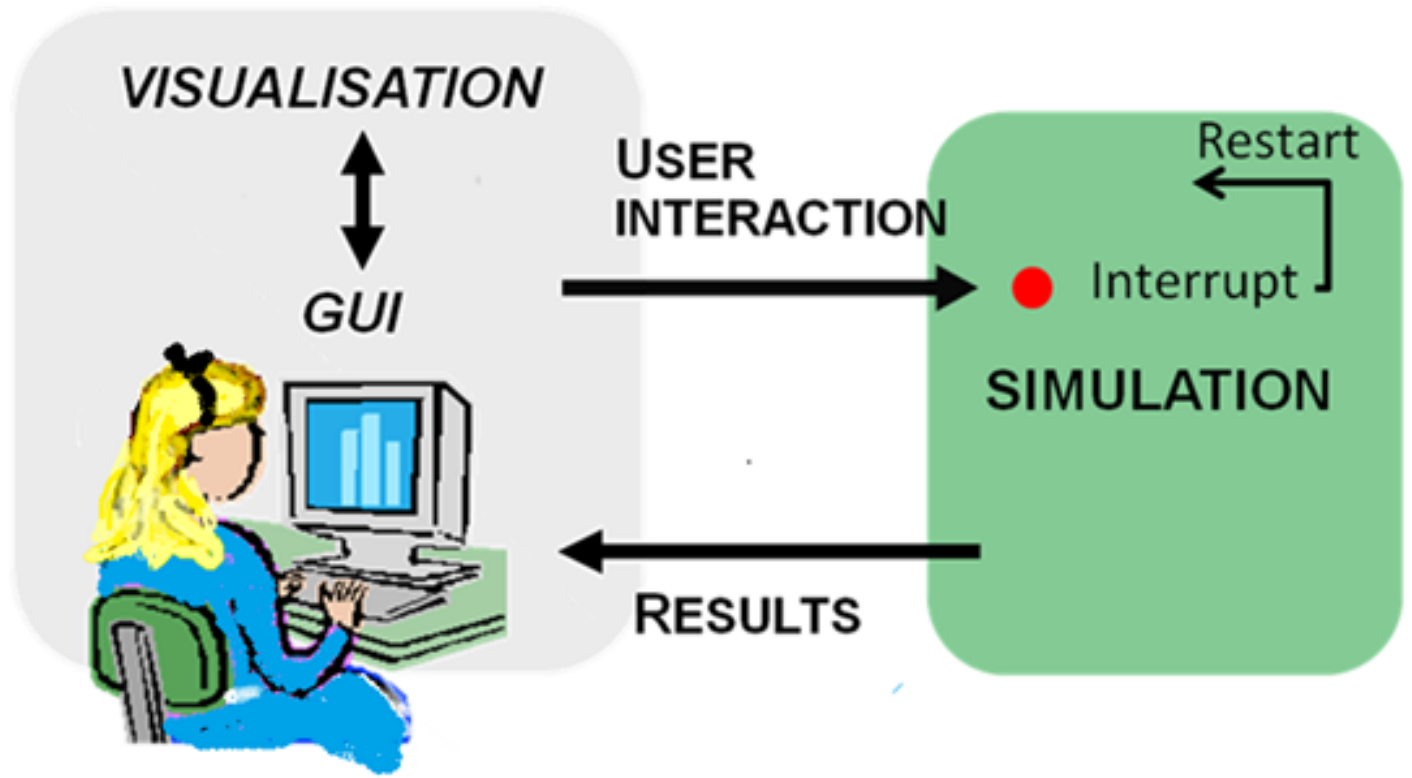}
\sidecaption 
\caption {A user guides an often time and memory consuming simulation in order
to build a solution to his/her problem via a graphical user interface.}
\label{fig:1}
\end{figure}

Interactive computing frameworks, libraries, and Problem Solving Environments
(PSEs) are used by specialists to interact with complex models, while not
requiring deep knowledge in algorithms, numerics, or visualisation techniques.
These are user-friendly facilities for guiding the numerically approximated
problem solution. The commonly agreed features are: a sophisticated user
interface for the visualisation of results on demand and a separated steerable,
often time- and memory-consuming simulation running on a high-performance
computer (see Fig.~\ref{fig:1}).

The concept has been present in the scientific and engineering community
already for more than two decades. Meanwhile, numerous powerful tools serving
this purpose have been developed. A brief overview of some state-of-the-art
tools -- steering environments and systems such as CSE~\cite{vanLiere.1996},
Magellan~\cite{Vetter.1997}, SCIRun~\cite{Parker.1998},
Uintah~\cite{deStGermain.2000}, G-HLAM~\cite{Rycerz.2006} and
EPSN~\cite{Nicolas.2007} , libraries such as CUMULVS~\cite{Geist.1997}, or
RealityGrid~\cite{Brooke.2003, Pickles.2004}, or frameworks such as
Steereo~\cite{Jenz.2010} -- is provided in the next section. Those tools differ
in the way they provide interactive access to the underlying simulation codes,
using check- and breakpoints, satellites connected to a data manager, or data
flow concepts, e.\,g., hence they cannot always fully exploit interactive
computing and are usually of limited scope concerning different application
domains.

\section{Computational Steering -- State of the art}
CSE~\cite{vanLiere.1996} is a computational steering environment consisting of
a very simple, flexible, minimalistic kernel and modular components, so-called
satellites, where all the higher level functionality is pushed. It is based on
the idea of a central process, i.\,e.\ a data manager to which all the
satellites can be connected. Satellites can create and read/write variables,
and they can subscribe to events such as notification of mutations of a
particular variable~\cite{Mulder.1995}. The data manager informs all the
satellites of changes made in the data and an interactive graphics editing tool
allows users to bind data variables to user interface elements.

CUMULVS~\cite{Geist.1997} is a library that provides steering functionality so
that a programmer can extract data from a running (possibly parallel)
simulation and send those data to the visualisation package. It encloses the
connection and data protocols needed to attach multiple visualisation and
steering components to a running application during execution. The user has to
declare in the application which parameters are allowed to be modified or
steered, or the rules for the decomposition of the parallel data, etc. Using
check-pointing, the simulation can be restarted according to the new settings.

In the steering system called Magellan~\cite{Vetter.1997}, steering objects are
exported from an application. A collection of instrumentation points, such as
so-called actuators, know how to change an object without disrupting
application execution. Pending update requests are stored in a shared buffer
until an application thread polls for them ~\cite{Vetter.1997}.

EPSN~\cite{Nicolas.2007} API is a distributed computational steering
environment, where an XML description of simulation scripts is introduced to
handle data and concurrency at instrumentation points. There is a simple
connection between the steering servers, i.\,e.\ simulation back ends, and
clients, i.\,e.\ user interfaces. When receiving requests, the server
determines their date, thus, the request is executed as soon as it fulfills a
condition. Reacting on a request means releasing the defined blocking points.

Steereo~\cite{Jenz.2010} is a light-weight steering framework, where the client
can send requests and the simulation will respond to them. However, the
requests are not processed immediately, but rather stored in a queue and
executed at predefined points in the simulation code. Hence, users have to
define when and how often this queue should be processed.

The RealityGrid~\cite{Brooke.2003, Pickles.2004} project has provided a highly
flexible and robust computing infrastructure for supporting the modelling of
complex systems \cite{RealityGrid.2003}. An application is structured into a
client, a simulation, and a visualisation unit communicating via calls to the
steering library functions. Also this infrastructure involves the insertion of
check- and break-points at fixed places in the code where changed parameters
are obtained and the simulation is to be restarted.

In the SCIRun~\cite{Parker.1998} problem solving environment (PSE) for
modelling, simulation, and visualisation of scientific problems, a user may
smoothly construct a network of required modules via a visual programming
interface. Computer simulations can then be executed, controlled, and tuned
interactively, triggering the re-execution only of the necessary modules, due
to the underlying dataflow model. It allows for extension to provide real-time
feedback even for large scale, long-running, data-intensive problems. This PSE
has typically been adopted to support pure thread-based parallel simulations so
far. Uintah~\cite{deStGermain.2000} is a component-based visual PSE that builds
upon the best features of the SCIRun PSE, specifically addressing the massively
parallel computations on petascale computing platforms.

In the G-HLAM~\cite{Rycerz.2006} PSE, the focus is more on fault tolerance,
i.\,e.\ monitoring and migration of the distributed federates. The group of
main G-HLAM services consists of one which coordinates management of the
simulation, one which decides when performance of a federate is not
satisfactory and migration is required, the other which stores information
about the location of local services. It uses distributed federations on the
Grid for the communication among simulation and visualisation components.

All of those powerful tools have, however, either limited scope of application,
or are involving major simulation code changes in order to be effective. This
was the motivation for us to design a new framework that incorporates the
strong aspects of the aforementioned tools, nevertheless overcomes their weak
aspects in order to provide a generic concept for a plenitude of different
applications with minimal codes changes and a maximum of interactivity.

Within the Chair for Computation in Engineering at Technische Universit\"{a}t
M\"{u}nchen, a series of successful Computational Steering research projects
took place in the previous decade. It has also involved collaboration with
industry partners. Performance analysis has been done for several interactive
applications, in regard to responsiveness to steering, and the factors limiting
performance have been identified. The focus at this time was on interactive
computational fluid dynamics (CFD), based on the Lattice-Boltzmann method,
including a Heating Ventilation Air-Conditioning (HVAC) system
simulator~\cite{Borrmann.2005}, online-CFD simulation of turbulent indoor flows
in CAD-generated virtual rooms~\cite{Wenisch.2004}, interactive thermal comfort
assessment~\cite{vanTreeck.2007}, and also on structure mechanics --
computational methods in orthopaedics. Over time, valuable observations and
experience have resulted in significant reduction of the work required to
extend an existing application code for steering.

Again, the developed concepts have been primarily adapted to this limited
number of application scenarios, thus, they allow for further investigations so
as to become more efficient, generic, and easy to implement. This is where our
framework comes into play.

\section{The Idea of the Framework}
For widening the scope of the steerable applications, an immediate response of
any simulation back end to the changes made by the user is required. Hence, the
regular course of the simulation has to be interrupted as soon as a user
interacts. Within our framework, we achieve this using the software equivalent
of hardware interrupts, i.\,e.\ signals. The check for updates is consequently
done in small, user-defined, cyclic intervals, i.\,e., within a function
handling the Unix ALARM signal.

If the check does not indicate any update from the user side, the simulation
gets the control back and continues from the state saved at the previous
interrupt-point. Otherwise, the new data is received, matched to the simulation
data (which is the responsibility of the user himself), and simulation state
variables (for instance loop delimiters) are manipulated in order to make the
computation stop and then automatically start anew according to the user
modifications. Taking the pseudo code of an iteratively executed function
(within several nested loops) as an example, the redundant computation is
skipped as soon as the end of the current, most-inner loop iteration is
reached. This is, namely, the earliest opportunity to compare the values of the
simulation state variables, and, if the result of the comparison indicates so,
exit all the loops (i.\,e.\ starting with most-inner one and finishing with the
most-outer one)~\cite{Knezevic.2011a, Knezevic.2011b, Knezevic.2010}. This
would exactly mean starting computation over again, as illustrated in the
pseudocode:

\begin{verbatim}
% X_end, Y_end declared global

begin function Signal_Handler()
  % manipulate X_end, Y_end so that the redundant
  % computation is skipped and started anew
  X_end = Y_end = -1  
end

% set time interval for periodically occurence of
% ALARM signal to stop execution and call handler
Set_Alarm()

% user function to be interrupted
begin function Compute()
  for t = T_start to T_end do  
    for i = X_start to X_end do
      for j = Y_start to Y_end do
        Process(data[i][j])
        % potential update is recognised next
      od                                
    od
  od
end
\end{verbatim}

As elaborated in~\cite{Knezevic.2010}, to guarantee the correct execution of a
program, one should use certain type qualifiers (provided by ANSI C, e.\,g.)
for the variables which are subject to sudden change or objects to interrupts.
One should ensure that certain types of objects which are being modified both
in the signal handler and the main computation are updated in an atomic way.
Furthermore, if the value in the signal handler has changed, the outdated value
in the register should not be used again. Instead, the new value should be
loaded from memory. This may occur due to the custom compiler optimisations. In
addition to this, sufficient steps have to be taken to prevent potentially
introduced severe memory leaks before the new computation is started. This is
due to the interrupts and their possible occurrence before the memory allocated
at a certain point has been released.

Finally, with either one or several number of iterations being finished without
an interrupt, new results are handed on to the user process for visualisation.
One more time it is the user's responsibility to prescribe to the front end
process how to interpret the received data so that it can be coherently
visualised~\cite{Knezevic.2011a, Knezevic.2011b, Knezevic.2010, Knezevic.2011c,
Knezevic.2011d}.

In Fortran, similar to C/C++, support for signal handling can be enabled at
user level with minimal efforts involved. Some vendor supplied Fortran
implementations, including for example Digital, IBM, Sun, and Intel, have the
extension that allows the user to do signal handling as in
C~\cite{Baloai.2008}. Here, a C wrapper function for overriding the default
signal behaviour has to be implemented. However, the behaviour of the Fortran
extension of the aforementioned function is implementation dependent, and if
the application is compiled using an Intel Fortran compiler, when the program
is interrupted, it will terminate unless one ``clears'' the previously defined
action first.

Due to the accuracy requirements and the increasing amount of data which has to
be handled in numerical simulations of complex physical phenomena nowadays,
there is an urge to fully exploit the general availability and increasing CPU
power of high-performance computers. For this, in addition to efficient
algorithms and data structures, sophisticated parallel programming methods are
a constraint. The design of our framework, therefore, takes into consideration
and supports different parallel paradigms, which results in an extra effort to
ensure correct program execution and avoid synchronisation problems when using
threads, as explained in the following subsection.

\subsection{Multithreading Parallelisation Scenario}
We consider the scenario when pure multithreading (with, e.\,g., OpenMP/POSIX
threads) is employed in the computations on the simulation side. Since a random
thread is interrupted via signal at the expiration of the user-specified
interval, that thread probes, via the functionality of the Message Passing
Interface (MPI), if any information regarding the user activity is available.
If the aforesaid checking for a user's message indicates that an update has
been sent, the receiving thread instantly obtains all the information and
applies necessary manipulations in order to re-start the computation with the
changed setting. Hence, all other threads also become instantly aware that
their computations should be started over again and must now proceed in a way
in which clean termination of the parallel region is guaranteed.

\subsection{``Hybrid'' Parallelisation Scenario}
\begin{figure}[h]
\includegraphics[scale=0.30]{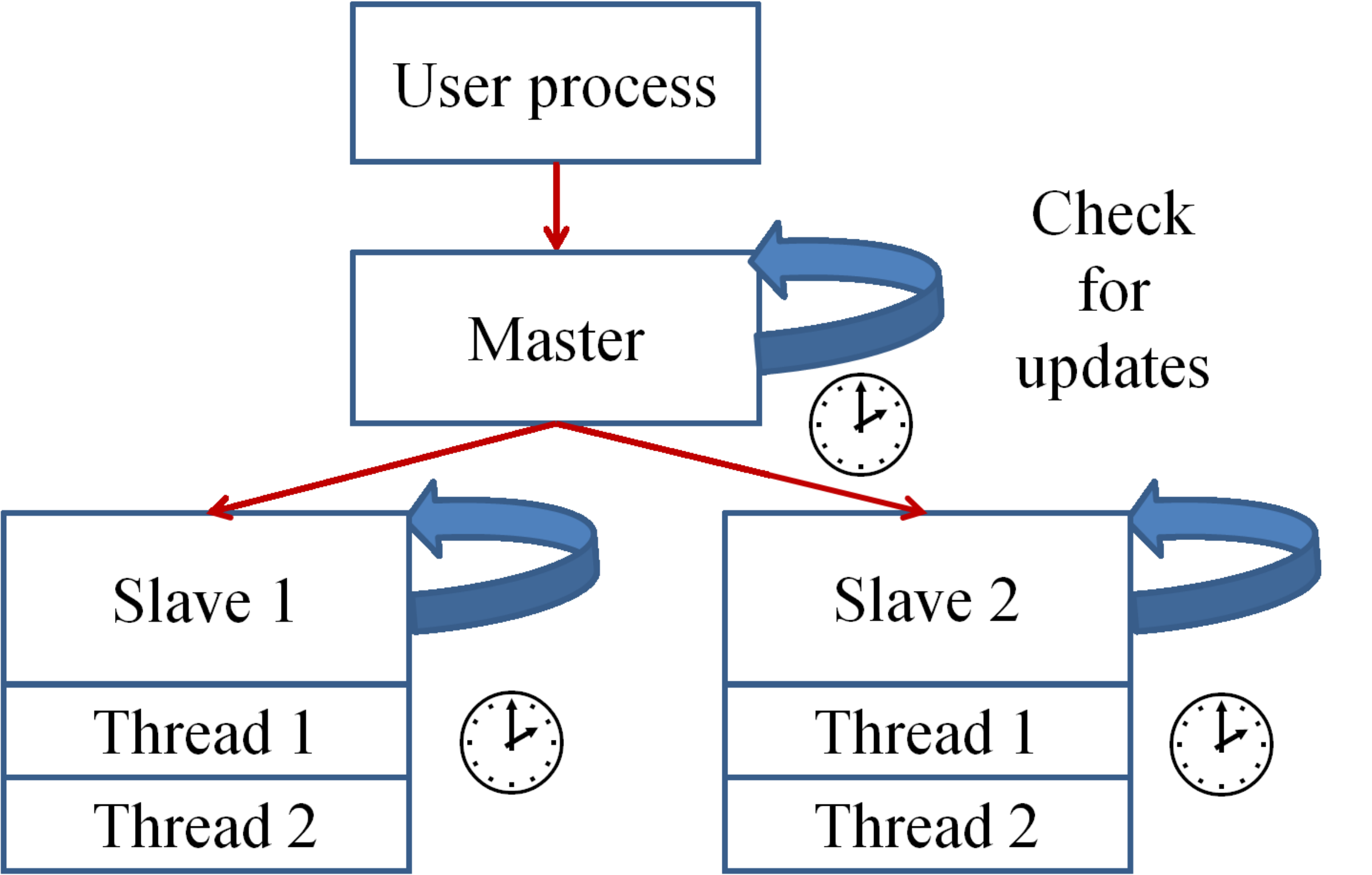}
\sidecaption 
\caption {In the case of a hybrid parallel scenario, each process is doing its
own checks for updates; one random thread per process is interrupted in small,
fixed time intervals.}
\label{fig:2}
\end{figure}

In a ``hybrid'' parallel scenario (i.\,e.\ MPI and OpenMP -- see
Fig.~\ref{fig:2}), a random thread in each active process is being interrupted,
hence, fetches an opportunity to check for the updates. The rest of the
procedure is similar as described for the pure multithreading, except that now
all the processes have to be explicitly notified about the changes performed by
a user. This may involve additional communication overheads. Moreover, if one
master process, which is the direct interface of the user's process to the
computing-nodes, i.\,e.\ slaves, is supposed to inform all of them about the
user updates, it may become a bottleneck. Therefore, a hierarchical
non-blocking broadcast algorithm for transferring the signal to all computing
nodes has been proposed in~\cite{Knezevic.2011a, Knezevic.2011b}.

\section{Applications}
In the following, a few application scenarios are presented, where the
implemented framework has been successfully integrated. First, a simple 2D
simulation of a temperature conduction, used only for testing purposes, where
heat sources, boundaries of the domain, etc.\ can be interactively modified.
Then, a neutron transport simulation developed at the Nuclear Engineering
Program, University of Utah, which has been the first Fortran test case for the
framework. The next one is the sophisticated Problem Solving Environment SCIRun
developed at the Scientific Computing and Imaging (SCI) Institute, University
of Utah. The final one is a tool for pre-operative planning of hip-joint
surgeries, done as a collaborative project of the Chair for Computation in
Engineering and Computer Graphics and Visualization at Technische
Universit\"{a}t M\"{u}nchen. A summary of necessary modifications of the
original codes in order to integrate the framework is discussed in
section~\ref{subsection:effort}.

\subsection{Test Case 1 -- A Simple Heat Conduction Simulation}

{\bf{Simulation:}} For proof of concept, we consider as a first, very simple
example a 2D simulation of heat conduction in a given region over time. It is
described by the Laplace equation, whose solutions are characterised by a
gradual smoothing of the starting temperature distribution by the heat flow
from warmer to colder areas of a domain. Thus, different states and initial
conditions will tend toward a stable equilibrium. After numerical treatment of
the PDE via a Finite Difference scheme, we come up with a five-point stencil.
The Gauss-Seidel iteration method is used to solve the resulting linear system
of equations.

{\bf{GUI/Visualisation:}} For interacting with the running simulation, a
graphical user interface is provided using the wxWidgets
library~\cite{wxWidgets}. The variations of the height along the z-axes,
pointing upward, are representing the variations of the temperature in the
corresponding 2D domain. Both the simulation and the visualisation are
implemented in C++ and are separate MPI processes.

\begin{figure}[h]
\includegraphics[scale=0.4]{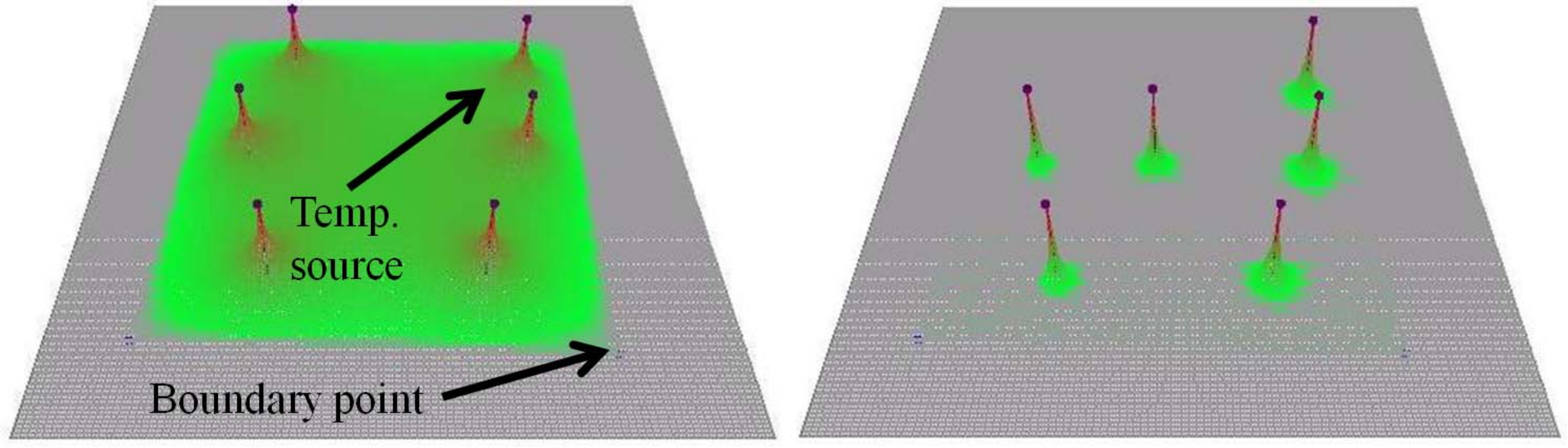}
\caption {Left: an initial scenario; right: moving heat sources/boundaries
leads to the restart of the computation, but in-between two too frequent
restarts a user is unable to estimate the equilibrium temperature in the region
farther away from the heat sources (here, further iterations of the solver
would be necessary).}
\label{fig:3}
\end{figure}

{\bf{User interaction:}} When it comes to the interplay with the program during
the simulation, there are a few possibilities available -- one can
interactively add, delete, or move heat sources, add, delete, or move boundary
points of the domain, or change the termination condition (maximal number of
iterations or error tolerance) of the solver. As soon as a user interacts, the
simulation becomes immediately aware of it and consequently the computation is
restarted. An instant estimation of the equilibrium state for points of the
domain far away from the heat sources is unfortunately not always feasible on
the finest grid used (300$\times$300). This may be the case due to the short
intervals between two restarts in case of too frequent user interaction, as
shown in Fig.~\ref{fig:3}. Here we profit from a hierarchical approach.

\begin{figure}[h]
\includegraphics[scale=0.3]{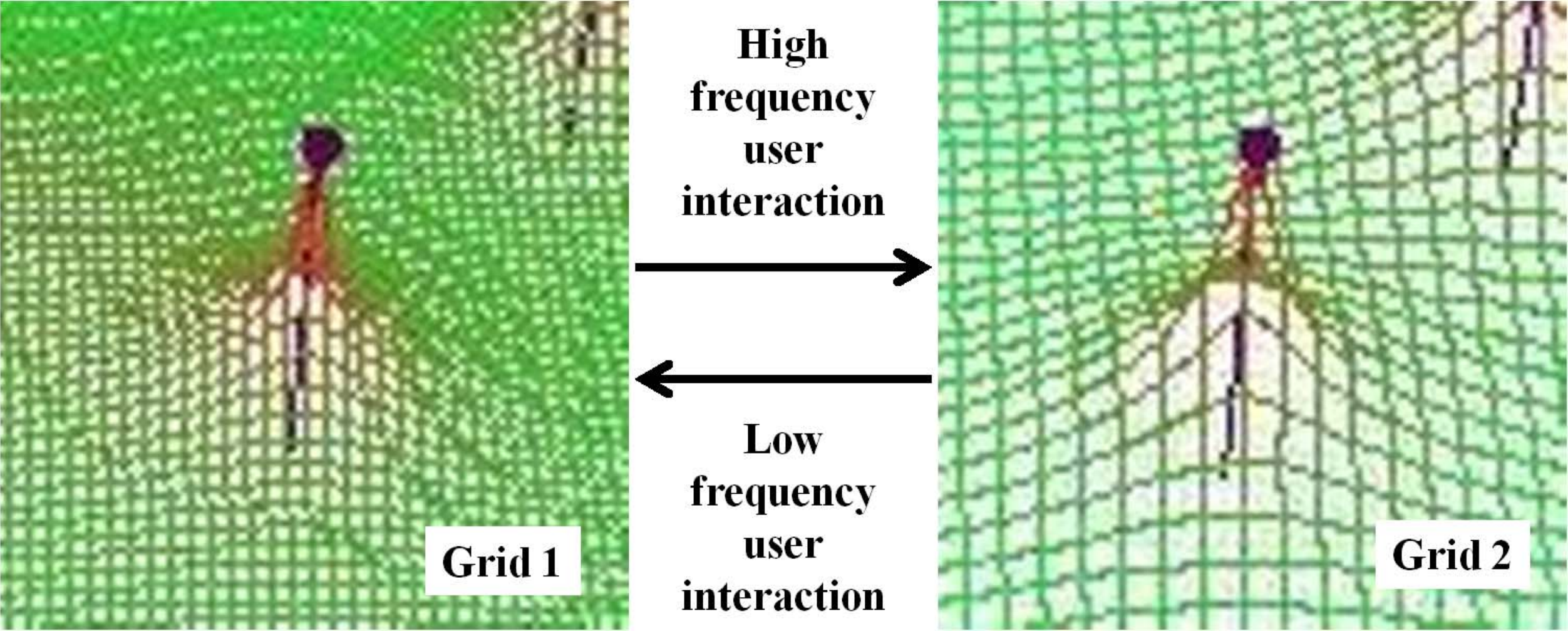}
\caption {Switching to a coarser grid in case of moving heat sources or
boundaries, switching back to the finer one once the user stops interacting.}
\label{fig:4}
\end{figure}

The {\bf{hierarchical approach}} is based on switching between several grids of
different resolutions depending on the frequency of the user interaction. At
the beginning, the finest desired grid is used for the computation. When the
simulation process is interrupted by an update, it restarts the computation
with the new settings, but on a coarser grid for faster feedback, i.\,e.\ to
provide new results as soon as possible. As long as the user is frequently
interacting, all computations are carried out on the coarser grids only. If the
user halts, i.\,e.\ stops interacting, the computation switches back to the
finest grid in order to provide more accurate values. In this particular test
case, three different grids were used -- an initial 300$\times$300 grid, the
four times smaller, intermediate one (150$\times$150, in case of lower pace of
interactions, i.\,e.\ adding/deleting heat sources or boundary points, e.\,g.)
and, finally, the coarsest one (75$\times$75) for very high frequency moving of
boundary points or heat sources over the domain (Fig.~\ref{fig:4}). The coarser
grids are not meant for obtaining quantitative solutions, just for a fast
qualitative idea how the solution might look like. If a user interactively
found an interesting setup, he just has to stop and an accurate solution for
this setup will be computed. Nevertheless, measurements concerning the
different grids showed that the variation of the solution on the finest grid
compared to the intermediate one is around 4.5\%, and compared to the coarsest
one around 14.6\%. The described approach, on the other hand, leads to an
improvement in convergence by a factor of 2.

Additionally, we employ a multi-level algorithm -- the results of the
computation on the coarsest grid are not discarded when switching to the finer
one. Our concept, namely, already involves a hierarchy of discretisations, as
is the case in multigrid algorithms, thus, we can profit from the analogous
idea. Our scheme is somewhat simpler -- it only starts with the solution on the
coarsest grid and uses the result we gain as an initial guess of the result on
a finer one. A set of examples has been tested (with grids
300$\times$300, 150$\times$150, and 75$\times$75) where the number of necessary
operations on the intermediate and fine grid could be halved. What seems to be
a somehow obvious approach, at least for this simple test scenario, can be
efficiently exploited in our forth test case where we use hierarchical Ansatz
functions for an interactive finite-element computation of a biomedical
problem.

In order to enable the framework functionality for interrupting the above
simulation, takes an experienced user a couple of hours at most. Implementing
the hierarchical approach (which is not a part of the framework) is more time
consuming (a few working days), since an optimal automatic detection when to
switch from one hierarchy to another has to be found---which requires numerous
experiments.

\subsection{Test Case 2 -- A Neutron Transport Simulation}
We present as second test case the integration of the framework into a
computationally efficient, high accuracy, geometry independent neutron
transport simulation. It makes researchers' and educators' interaction with
virtual models of nuclear reactors or their parts possible.

{\bf{Simulation:}} AGENT (Arbitrary GEometry Neutron Transport) solves the
Boltzmann transport equation, both in 2D and 3D, using the Method of
Characteristics (MOC)~\cite{Lee.2004}. The motivation for steering such a
simulation during runtime comes mostly from the geometric limitation of this
method, which requires fine spatial discretisation in order to provide an
accurate solution to the problem. On the other hand, a good initial solution
guess would help tremendously to speed-up the convergence, and this property is
used to profit from our framework. The 3D discretisation basis for the
Boltzmann equation consists of the discrete number of plains, for each of which
both a regular geometry mesh and a number of parallel rays in a discrete number
of directions are generated. The approximation results in a system of equations
to be iteratively solved for discrete fluxes.

{\bf{GUI/Visualisation:}} The result in terms of the scalar fluxes is
simultaneously calculated and periodically visualised. The simulation server
maintains a list of available simulation states, and clients connect using the
ImageVis3D volume rendering tool~\cite{Fogal.2010} to visualise the results in
real time. Users can interfere with the running simulation via a simple console
interface, providing the new values of the desired parameters.

{\bf{User interaction:}} Instant response of the simulation to the changes made
by the user is again achieved via signals. Using the technique described as our
general concept, the most outer iteration instantly starts anew, as soon as its
overall state is reset within the main computational steering loop, according
to the updated settings and necessary re-initialisation of the data. By
manipulating only two simulation parameters in the signal handler, it is
achieved that the iteration restarts almost within a second in all the cases --
e.\,g.\ 20 planes in z-direction, each discretised by a 300$\times$300 grid and
36 azimuthal angles (where only one, the most outer iteration lasts
approximately 500 seconds). The effort to integrate our framework into this
application depends on whether the re-allocation of the memory and
re-initialisation of the data is required, and if one wants to re-use the
values from the previous iterations~\cite{Knezevic.2012a, Knezevic.2012b}.

\begin{figure}[h]
\includegraphics[scale=0.5]{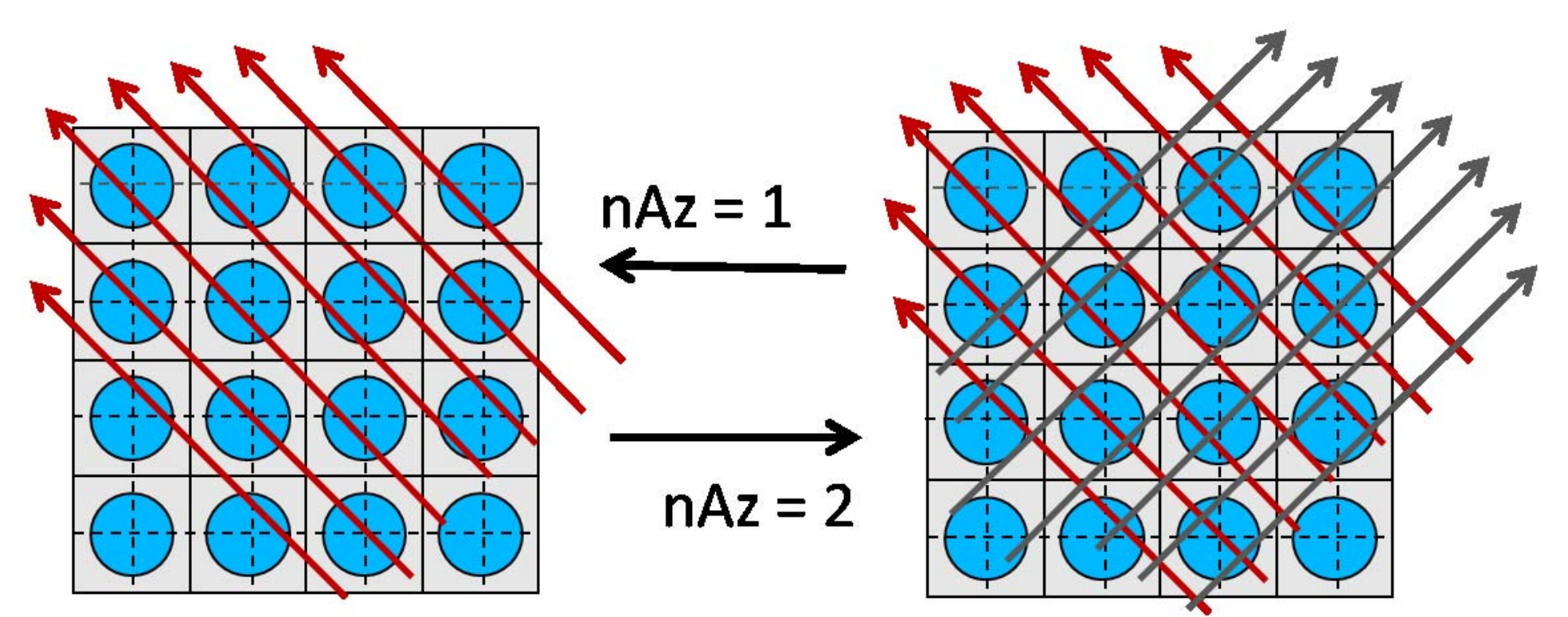}
\caption {Experimenting with different numbers of azimuthal angles, small
values are given to simplify the picture.}
\label{fig:5}
\end{figure}

{\bf{Hierarchical and multilevel approach:}} It is likely, similar to the heat
conduction scenario, that the user wants to accelerate the convergence by
starting calculations with lower accuracy (i.\,e.\ number of azimuthal angles,
see Fig.~\ref{fig:5}), preserve and re-use some of the values from the previous
calculation as an initial guess for the higher accuracy solution. For a
conceptually similar algorithm, such as the previously described multilevel
approach in the 2D heat conduction simulation, we have seen that our framework
has given promising results. The re-initialisation of the data for this, most
challenging, scenario is a part of imminent research.

To briefly conclude on this application scenario, the integration of the
framework has been straightforward and also not very time consuming. After
examining the initial code, deciding which variables to register within the
framework, and writing reinitialisation routines, it has taken a few hours to
couple the components together and enable visualisation after each iteration.
The major effort refers actually to the re-initialisation of variables at the
beginning of each ``new'' computation, i.\,e.\ after a user interaction, which
is also not a responsibility of the framework itself.

\subsection{Test Case 3 -- Extension of a Problem Solving Environment}
{\bf{Simulation:}} As mentioned before, SCIRun is a PSE intended for
interactive construction, debugging, and steering of large-scale, typically
parallel, scientific computations~\cite{Shepherd.2009}. SCIRun simulations are
designed as networks of computational components, i.\,e.\ modules connected via
input/output ports. This makes it very easy for a programmer to modify a module
without affecting others. As SCIRun is already a mature, sophisticated
environment for computational steering, nevertheless, our goal is to improve it
in a way that real time feedback for extensive time- and memory-consuming
simulations becomes possible. Here, SCIRun needs to finish an update first
before new results are shown, which easily can lead to long latencies between
cause and effect.

{\bf{GUI/Visualisation:}} For the user, it is possible to view intermediate
results after a pre-defined number of iterations, while the calculations
continue to progress. At some point, he may require to influence the current
simulation setup. Different options such as parameter modification for each
module are available via corresponding interfaces. Both the modified module and
modules whose input data is dependent on that module's output are stored in a
queue for execution. Our intention is to interrupt the module currently being
executed and skip the redundant cycles, as well as to remove any module
previously scheduled for execution from the actual queue.

{\bf{User Interaction:}} The concept has been tested on several examples to
evaluate the simulation response to the modifications during runtime. These
scenarios are: a simulation that facilitates early detection of acute heart
ischemia and two defibrillation-like simulations -- one on a homogeneous cube
and the other on a human torso domain. The challenges of getting an immediate
feedback/response of the simulation depend on a few factors -- the size of the
problem, the choice of the modified parameters within the simulation, etc. The
earlier in the execution pipeline the parameter appears, the more modules have
to be re-executed, thus, the more challenging it is to provide the real-time
response to the user changes. A user can define different discretisation
parameters for a FEM computation such as the mesh resolution for all spatial
directions. For solving the resulting linear system of equations, different
iterative solvers as well as pre-conditioners can be used; one may change
tolerances, the maximal number of iterations, levels of accuracy, as well as
other numerical or some more simulation-specific parameters. In the created
network of modules, typically the most laborious step is the SolveLinearSystem
module. Thus, the first challenge is how to interrupt it as soon as any change
is made by the user -- in particular, the changes done via UI to this module.
To achieve this in the algorithm of the linear equation solver, the maximal
number of iterations (a user interface variable) is manipulated in the signal
handler, so as to be set to some value outside of the domain of the iterator
index which interrupts the simulation as described before. The execute function
of this module also has to be re-scheduled afterward with the new user-applied
settings. However, one has to take care that the previous interrupted execution
of the same module is finished in a clean way and that the execute function has
to be called anew (in order to trigger re-computation instantly). If one
chooses to emit the partial solution after each iteration, executions of
several visualisation modules are scheduled after each iteration, which would
take additional few seconds after each iteration. This is because after an
interrupted iteration the preview of old results has to be cancelled. The
execution of all modules, which would happen after SolveLinearSystem, has to be
aborted. The scheduler cancels the execution of all the scheduled modules that
have not begun yet by making sure an exception is employed. Changing any input
field of a module via its UI automatically triggers the re-execution of all the
modules following it in the pipeline.

\subsubsection{Tool for early detection of heart ischemia}
Myocardial ischemia is characterised by reduced blood supply of the heart
muscle, usually due to coronary artery disease. It is the most common cause of
death in most Western countries, and a major cause of hospital admissions
~\cite{Podrid.2005}. By early detection further complications might be
prevented. The aim of this application is the generation of a quasi-static
volume conductor model of an ischemic heart, based on data from actual
experiments ~\cite{Stinstra.2012}. The generation of models of the myocardium
is based on MR images/scans of a dog heart. The known values are extracellular
cardiac potentials as measured by electrodes on an isolated heart or with
inserted needles. The potential difference between the intracellular and
extracellular space which is being calculated is not the same for ischemic and
healthy cells. A network of modules is constructed within SCIRun to simulate
and then render a model of the transmembrane potential of a dog's myocardium in
experiments (Fig.~\ref{fig:6}).

\subsubsection{Defibrillation}
\begin{figure}[h]
\includegraphics[scale=0.4]{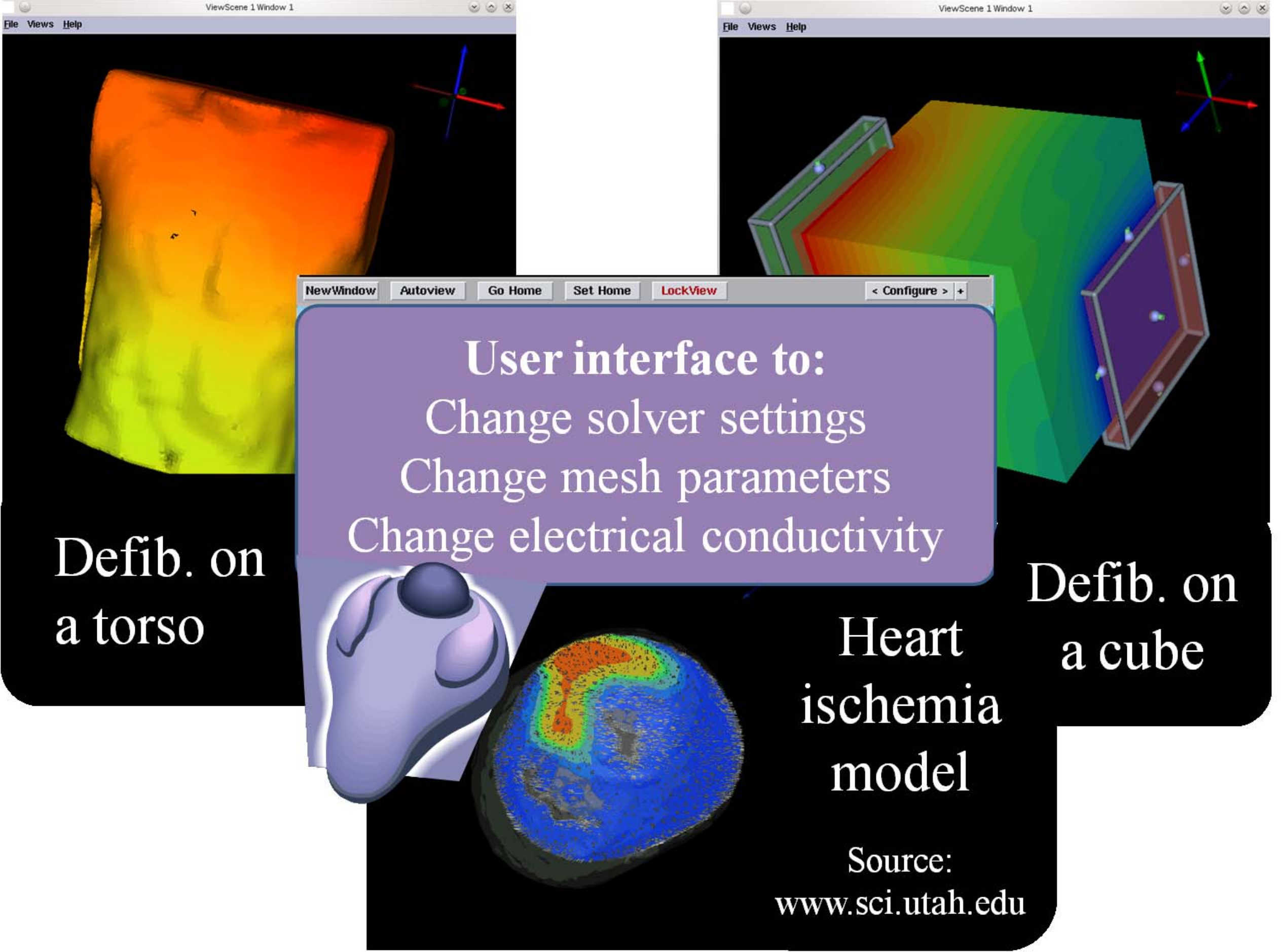}
\caption {Illustrated user interfaces for the tested simulation scenario.}
\label{fig:6}
\end{figure}

Defibrillation therapy consists of delivering a dose of electrical energy to
the heart with a device that terminates the arrhythmia and allows normal sinus
rhythm to be re-established by the body's natural pacemaker. Implantable
Cardioverter Defibrillators (ICDs) are relatively common, patient specific,
implantable devices that provide an electric shock to treat fatal arrhythmias
in cardiac patients ~\cite{Steffen.2012}. By building a computational model of
a patient's body with ICDs and mapping conductivity values over the entire
domain, we can accurately compute how activity generated in one region would be
remotely measured in another region ~\cite{Weinstein.2005}, which is exactly
what doctors would be interested in. First, we consider a simulation of the
electrical conduction on a homogeneous cube domain (Fig.~\ref{fig:6}) with two
electrodes placed within. Each of the electrodes is assigned a conductivity
value. The effect of changing those values is explored for both of the
electrodes. The second example helps to determine optimal energy discharge and
placement of the ICD in the human torso (Fig.~\ref{fig:6}). A model of the
torso into which ICD geometry is interactively placed is based on patient MRI
or CT data. Different solver-related parameters for the resulting system of the
linear equations, conductivity values, as well as mesh resolutions for a FEM
computation can be applied during runtime. This allows for previewing the
solution on a coarser grid and switching to finer ones, once the user is
satisfied with the current setting.

For a user to integrate the framework, the major effort has been related to
re-triggering the execution of all the needed modules when the user makes a
change. This has required good understanding of a used Model-View-Controller
pattern. On the other hand, registering the variables which need to be
manipulated within the framework to interrupt the execution of the modules of
interest, has required negligible amount of time.

\subsection{Test Case 4 -- A Biomedical Application}
Another test case is an analysis tool which assists an orthopaedic surgeon to
do optimal implant selection and positioning based on prediction of response of
a patient-specific bone (femur) to a load that is applied. The tool consists of
two coupled components.

{\bf{Simulation:}} The first one is a simulation core, where the generated
models of femur geometry are based on CT/MRI-data and the computation is done
using the Finite Cell Method (FCM). FCM is a variant of high order
\emph{p}-FEM, i.\,e.\ convergence is achieved by increasing the polynomial
degree \emph{p} of the Ansatz functions on a fixed mesh instead of decreasing
the mesh sizes \emph{h} as in case of classical \emph{h}-FEM, with a fictitious
domain approach, as proposed in~\cite{Duester.2009}. With this method, models
with complicated geometries or multiple material interfaces can be easily
handled without an explicit 3D mesh generation. This is especially advantageous
for interactive computational steering, where this typically user-interaction
intensive step would have to be re-executed for each new configuration.

{\bf{GUI/Visualisation}} The second component is a sophisticated visualisation
and user interface platform that allows the intuitive exploration of the bone
geometry and its mechanical response to applied loads in the physiological and
the post-operative state of an implant-bone in terms of stresses and strains
\cite{Dick.2008, Dick.2009}. Thus, after updating the settings -- either after
insertion/moving an implant, or testing a new position/magnitude of the forces
applied to the bone -- for each unknown a scalar value, i.\,e.\ the so-called
von Mises stress norm, can be calculated and the overall result sent to the
front end to be visualised.

Some of the challenges in developing such an analysis tool are described in
more detail in~\cite{Yang.2010, Dick.2009, Dick.2008}. We conveniently had the
described simulation and a sophisticated user interface with visualisation
module as a starting point. Due to the initial rigid communication pattern
between the two components, however, a new setting could be recognised within
the simulation only after the results for the previous, outdated, one have been
completely calculated and sent to the user. Consequently, the higher polynomial
degrees \emph{p} were used, the dramatically longer became the total time until
one could finally perceive the effect of his last change. The integration of
our framework then comes into play not only to make the way the data is
communicated more suitable for this purpose, but also to enable interrupting
the simulation immediately and getting instant feedback ensued by any user
interaction.

For the best performance, on the front end, the main thread (in charge of
fetching user interaction data and continuous rendering), the second thread (in
charge of collecting and sending updates in timely fashion, via non-blocking
MPI routines), and the third thread (dedicated for waiting to receive results
as soon as these are available), are not synchronised with one another. This
way, we tackle the problem of long delays that would occur if one thread is
responsible for everything and communication is blocked as long as the thread
is busy, which would hinder the user in (smoothly) exploring the effects of his
interaction.

On the simulation side, as mentioned before, a variant of FEM is used.
Mainstream approaches are
\begin{itemize}
\item \emph{h}-FEM: convergence due to smaller diameters \emph{h} of elements,
\item \emph{p}-FEM: convergence due to higher polynomial degrees \emph{p},
\item \emph{hp}-FEM: combining the aformentioned ones by alternating \emph{h}
and \emph{p} refinements,
\item \emph{rp}-FEM: a combination of mesh repositioning and \emph{p}
refinements,
\item $\ldots$
\end{itemize}
In our case, for the algebraic equations gained by the \emph{p}-version Finite
Element Method describing the behaviour of the femur, iterative solvers such as
CG or multi grid could not be efficiently deployed due to the poor condition
number of the system. To make most out of the simulation performance potential,
a hierarchical concept based on an octree-decomposition of the domain in
combination with a nested dissection solver is used~\cite{Mundani.2007}. It
allows for both the design of sophisticated steerable solvers as well as for
advanced parallelisation strategies, both of which are indispensable within
interactive applications.

\begin{figure}[h]
\includegraphics[scale=0.4]{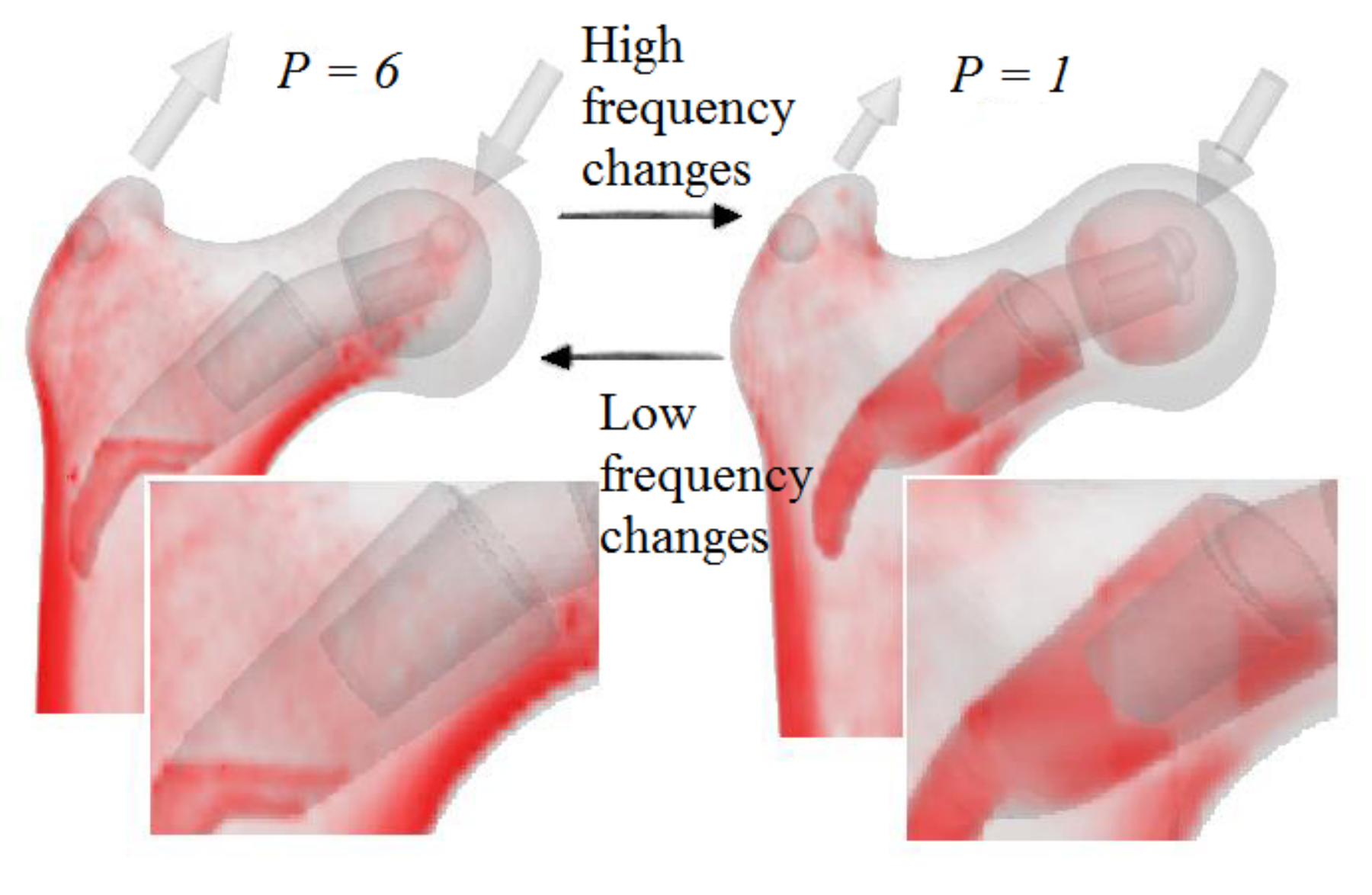}
\sidecaption 
\caption {Direct transition from $p = 6$ to $p = 1$ as soon as the user changes
the force's magnitude and direction, inserts an implant or moves it, while
gradually increasing from $p = 1 \rightarrow p = 2 \rightarrow p = 4
\rightarrow \ldots$ as soon as the user diminishes or finally stops his
interaction. Hence, a qualitative feedback about stress distribution for $p =
1$ or $p = 2$ is received instantly, finer result for $p \geq 4$ on demand.}
\label{fig:7}
\end{figure}

{\bf{User interaction:}} By applying a nested dissection solver, the most time
consuming step is the recursive assembly of the stiffness matrices, each
corresponding to one tree node, traversing the octree bottom up. Again,
cyclically-repeating signals are used for frequent checks for updates. If there
is an indicator of an upcoming message from the user side, this is recognised
while processing one of the tree nodes and the simulation variables are set in
a way which ensures skipping the rest of them. All the recursive assembly
function calls return immediately, and the new data is received in the next
step of the interactive computing loop (updating one or more of the leaf
nodes). Here, precious time has been saved by skipping all the redundant
calculations and, thus, calculating results only for an actual setting. As soon
as the whole assembly has been completed without an user interrupt, the result
in terms of stresses is sent back to the front end process for visual display.
However, there is an unavoidable delay of any visual feedback especially for
higher \emph{p} values, i.\,e.\ $p > 4$, in case of the used hardware and the
complexity of the geometric model. Namely, the time needed for a (full) new
computation is dramatically increasing in case of increasing \emph{p}. Thus, we
profit from a hierarchical approach one more time. The hierarchy exploited in
this approach refers to the usage of several different polynomial degrees
chosen by the user (Fig.~\ref{fig:7}). While the user's interplay with the
simulation is very intensive, he retrieves immediate feedback concerning the
effects of his changes for lower \emph{p}, being able to see more accurate
results (for higher \emph{p}) as soon as he stops interacting and let one
iteration finish. In this case, the computation is gradually switched to higher
levels of hierarchy, i.\,e.\ from $p=1$ to $p=2$ to $p=4$ and so on. The number
of MPI program instances, being executed in parallel for different \emph{p} can
be chosen by the user. A detailed communication schemes can be found in
\cite{Knezevic.2011c, Knezevic.2011d}.

To get several updates per second even for higher \emph{p} values, one has to
employ sophisticated parallelisation strategies. Custom decomposition
techniques (i.\,e.\ recursive bisection) in this scenario, as in case of long
structures such as a femur, typically hinder the efficient exploitation of the
underlying computing power as this leads to improper load distributions due to
large separators within the nested dissection approach. Thus, our next goal has
been the development of an efficient load balancing strategy for the existing
structural simulation of the bone stresses.
 
Task scheduling strategies typically involve a trade-off between the uniform
work load distribution among all processors, as well as keeping both the
communication and optimisation costs minimal. For hierarchically organised
tasks with bottom-up dependencies, such as in our generated octree structure,
the number of processors participating in the computation decreases by a factor
of eight in each level, similar to the problem posed by Minsky for the parallel
summation of $2N$ numbers with $N$ processors in a binary tree
\cite{Knezevic.2012c}.

In interactive applications which assume the aforementioned frequent updates
from user's side, those rapid changes within the simulation and tasks' state
favour static in comparison to dynamic load balancing strategies. It would also
have to be taken into consideration that certain modifications performed by a
user may involve major changes of the computational model. In this case, for
repeatedly achieving the optimal amount of work being assigned to each process
for each new user update, the overhead-prone scheduling step has to be executed
each time. Therefore, an efficient, nevertheless simple to compute scheduling
optimisation approach is needed.

\begin{figure}[h]
\includegraphics[scale=0.5]{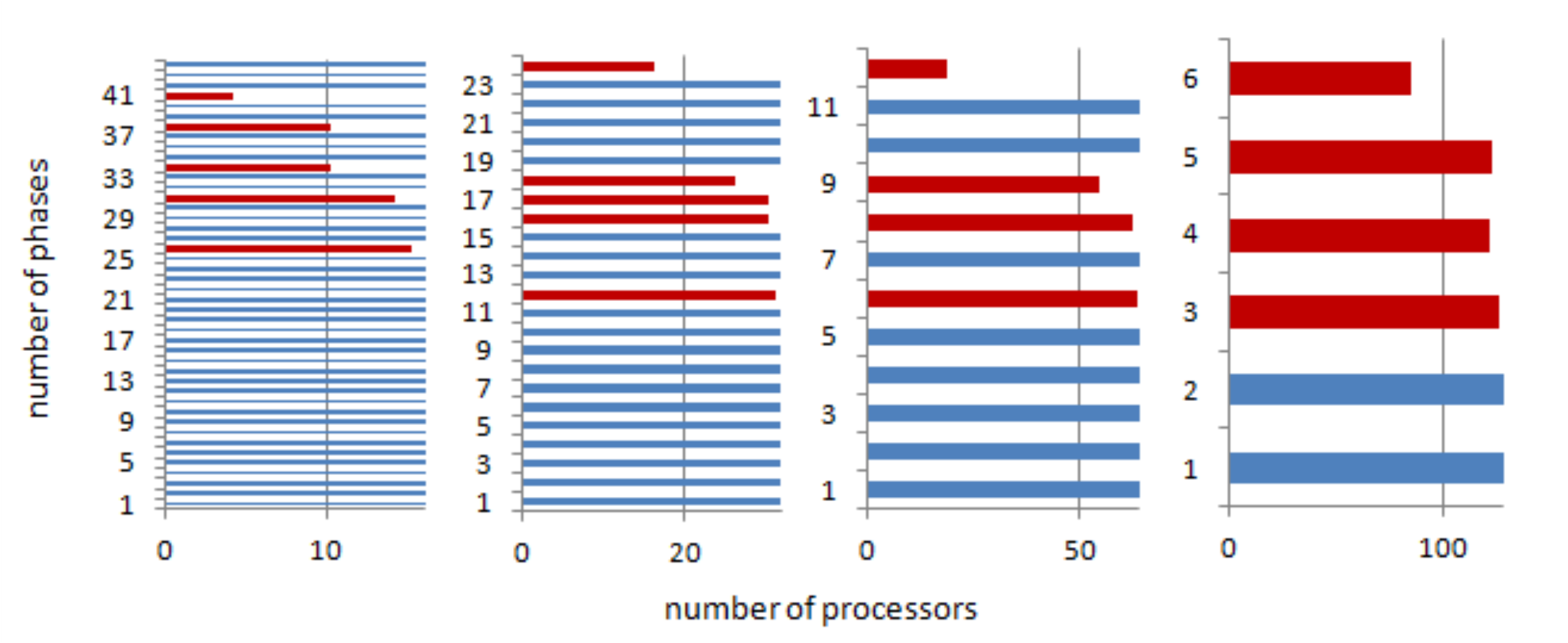}
\caption {Vertical axes describe the so-called ``phases'' and the horizontal
axes the number of processors involved in the particular phase. One ``phase"
involves actually the processors to which a task is assigned at that point.
Having the capacity of each phase as \emph{full} as possible is achieved,
i.\,e.\ all processors are busy with the approximately equal amount of work
throughout the solver execution.}
\label{fig:8}
\end{figure}

Since the scheduling problem can be solved by polynomial-depth backtrack
search, thus, is \emph{NP} complete for most of its variants, efficient
heuristics have to be devised. In our case, the sizes of the tasks, as well as
the dependencies among them (given by the octree structure responsible for the
order of the nested dissection advance) have to be considered. By making
decisions, we consider (1) the level of the task dependency in the tree
hierarchy where children nodes have to be processed before their parent nodes;
(2) among equal tasks (i.\,e.\ of the same dependency level) we distinguish
between different levels in the tree hierarchy, calling this property the
processing order. If the depth of the tree is \emph{H}, tasks from level
\emph{M} in the tree hierarchy have the processing order of $H - M - 1$. Then
we form lists of priorities, based on these two criteria, since tasks inside
very long branches of the tree with an estimated bigger load should be given a
higher priority. Additionally, we resort to a so-called max-min order, making
sure that big tasks, in terms of their estimated number of floating-point
operations, are the first ones assigned to the processors. We also split a
single task among several processors when mapping tasks to processors, based on
the comparison of a task's estimated work with a pre-defined `unit' task. This
way, arrays of tasks, so-called `phases', are formed, each phase consisting of
as many generated tasks as there are computing resources. Namely, taken from
the priority lists, tasks are assigned to phases in round-robin manner. The
results are illustrated in Fig.~\ref{fig:8}.

Those phases refer to the mapping which will be done during runtime of the
simulation. When the tasks are statically assigned to the processors, all of
them execute the required computations, communicating the data when needed and
also taking care that the communication delays due to the MPI internal
decisions are avoided, as elaborated more in \cite{Knezevic.2012c}.

Satisfactory speedup is achieved for different polynomial degrees \emph{p}
within the FCM computation, where higher polynomial degrees correspond to more
unknowns. Tests are being done currently for a larger number of distributed
memory computational resources. According to the tendency observed for up to 7
processors so far, engagement of larger numbers of processes would result in
the desired rate of at least several updates per second (i.\,e.\ 1--10\,Hz) for
the calculated bone stresses even for $p = 4$ or $p = 6$.

Referring back to the existing environment, without the integration of the
developed distributed parallel solver, the major effor that has been invested
in creating the new communication pattern to support the described hierarchical
approach was in the order of several working day. Anyhow, the functionality for
interrupting the computation to do checks for updates, thus, start a
computation anew if needed has been quick and straightforward.

\section{Results and Conclusions}
Finally, after discussing the achievements concerning interaction for each
application scenario in the previous section, here results in terms of
execution time overhead after integrating the framework in different scenarios
are to be presented, as well as the coding effort to be invested when
integrating the framework into an existing application code. Furthermore,
conclusions concerning the proposed hierarchical approaches are made and
possible ideas for further extension of the framework are discussed.

\subsection{Overhead of the Framework}
For the heat conduction application scenario, the integration of our framework
resulted in not more than 5--10\,\% overhead in the execution time. Tests have
been done also for the same problem with a message-passing-based parallel
Jacobi solver. Not even in case when user interaction was invoked in
5-millisecond intervals (which is far more frequent than it typically occurs in
practice) any significant effect of the interrupts on the overall execution
time (less than 10\,\%) was to be observed.

Performance evaluation of the biomedical test scenario, where the simulation is
executed on a multi-core architecture and connected to a visualisation front
end via a network, still proved that the overhead caused by the framework
itself is not significant (up to 11.7\,\%).

We have also tested the different simulation scenarios from SCIRun. The
measurements have been made for different update intervals, namely, 5, 2, or 1
millisecond for different solvers of linear systems of equations. In one of the
test case scenarios, for the shortest interval (i.\,e.\ 1 millisecond), the
overhead caused by the framework was up to 15\,\%. However, by making the
intervals longer (2 or 5 milliseconds, e.\,g.), the overhead was reduced to
5\,\% and 3\,\%, resp. When increasing the interval up to 5 milliseconds (and
beyond), an end-user does not observe the difference in terms of simulation
response. Hence, it is always recommendable to experiment with different
intervals for a specific simulation.

\begin{figure}
\includegraphics[scale=0.75]{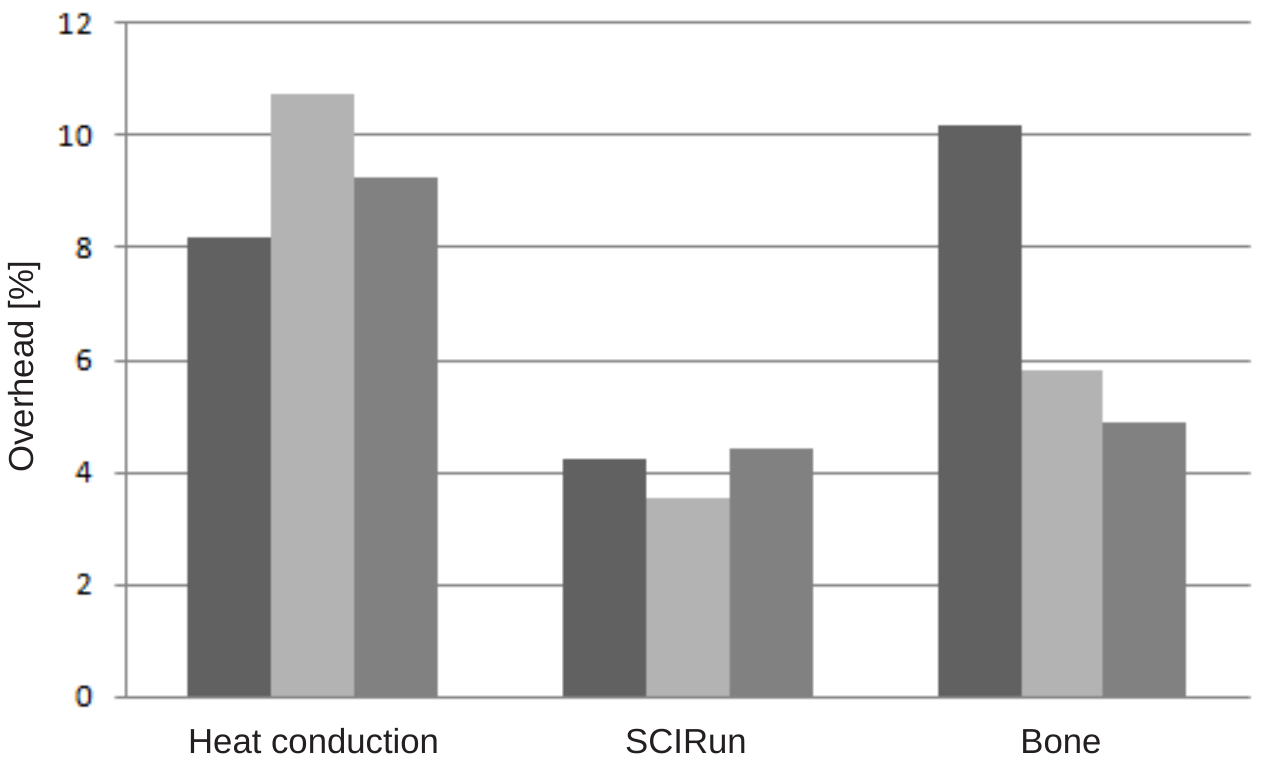}
\caption{Performance measurements: overhead of the framework (expressed in
terms of additional execution time) for alarm set to 1 millisecond -- heat
conduction simulation ($300\times300$ grid), executed on 1, 2, and 4 cores
(left to right); SCIRun PSE, heart ischemia example using CG, BCG, and MINRES
solver (left to right); biomedical application ($p = 4$), executed on 1, 2, and
4 cores (left to right).}
\label{fig:overhead}
\end{figure}

Some of the measurements are illustrated in Fig.~\ref{fig:overhead} for
comparison.

\subsection{User Effort for Integrating the Framework}
\label{subsection:effort}
A few modifications within any application code have to be made by the user in
order to integrate our framework. These modifications are -- as intended --
only minor, hence, we list all of them. All variables which will be affected by
the interrupt handler in order to force the restart of the computation have to
be declared global (to become visible in a signal handler). It is typically
enough to have only few of them, such as loop delimiters, in order to skip all
the redundant computations. If these variables shall be used also in the rest
of the code, a user can rename those he wants to manipulate within the signal
handler and declare only those as global. Atomicity of data updates and
prevention of compiler optimisations -- which would lead to incorrect value
references -- have to be ensured. The integrity of each user-defined `atomic'
sequence of instructions in the simulation code has to be provided. The calls
to the appropriate send and receive functions which are interface to our
framework have to be included in the appropriate places in the code. The user
himself should provide the correct interpretation of the data (in the receive
buffers of both simulation and visualisation components). Finally, he has to
enable the regular checks for updates by including appropriate functions which
will examine and change the default signal (interrupt) action, specifying the
time interval in which the checks of the simulation process(es) are made, as
shown in the following pseudo code example.

\begin{verbatim}
% Function to override the default signal action
begin func My_sig_action () 
  if update_available then
    receive update
    manipulate simulation specific variables
  fi
end

% Declare simulation specific variables to be
% global, atomic, and volatile 
begin func main ()
  Set_sig_action (My_sig_action)
  Set_interrupt_interval (time_slot)
end
\end{verbatim}

\subsection{Hierarchical Approaches} 
As one may also conclude, no matter how generic our basic idea is, when
applying it to the wide diversity of applications, the user himself has to be
involved in making certain decisions. For example, in our first test case, he
has to specify the number of grids which he would like to use together with
their resolutions. This information might be based on his previous experience,
i.\,e.\ at which resolution the problem can be solved within less than a second
(for choosing the coarsest grid), etc. The hierarchical approaches used so far
should not be the limitation for future test cases. In addition to recursively
coarsening the grid, or increasing the resolution of other simulation-specific
discretisations such as the number of azimuthal angles in AGENT, or increasing
the polynomial degree \emph{p} in the biomedical example, one may analogously
profit from his or her own simulation-specific hierarchical structures. Any
user of the framework can, if needed, easily adopt it to his individual
requirements.

\subsection{Outlook} 
In the future, we would like to tackle the computational expensive scenarios
with massively parallel simulations. In efforts to interrupt one thread per
process, a trade-off between ensuring a minimal number of checks per process
and allowing for receiving the data promptly is to be faced. Thus, an optimal
interval between the interrupts on different levels of the communication
hierarchy is going to be estimated. In addition, a possibility of distributing
the tasks among several user processes, each in charge of a certain group of
simulation processes will be examined to avoid typical master-slave
bottlenecks. Furthermore, we would like to explore techniques for the fast
transfer of (distributed) simulation results between front and back end,
especially in case of huge data sets, needed for an interactive visualisation.

\begin{acknowledgement}
The overall work has been financially supported by the Munich Centre of
Advanced Computing (MAC) and the International Graduate School of Science and
Engineering (IGSSE) at Technische Universit\"{a}t M\"{u}nchen and we would like
to gratefully acknowledge that. The work related to SCIRun PSE was made
possible in part by software from the NIH/NIGMS Center for Integrative
Biomedical Computing, 2P41 RR0112553-12. It was accomplished in winter 2011/12
during a three-month research visit of Jovana Kne\v{z}evi\'{c} to the
Scientific Computing and Imaging (SCI) Institute, University of Utah. She would
like to express her appreciation and gratitude to Prof.~Chris Johnson for
inviting her and all the researchers for fruitful discussions. Furthermore, she
would like to thank Hermilo Hern\'{a}ndez and Tatjana Jevremovi\'{c} at Nuclear
Engineering Program, University of Utah, and Thomas Fogal from SCI Institute,
in collaboration with whom the work on the AGENT project was done.
\end{acknowledgement}

\bibliographystyle{unsrt}
\bibliography{paper}

\end{document}